\documentclass[12pt]{article}

\usepackage[english]{babel}

\usepackage{amsmath}
\usepackage{graphicx}
\usepackage[colorlinks=true, allcolors=blue]{hyperref}

\newcommand{\that}{\widehat{\theta}}
\newcommand{\combi}[2]{\left(
  \begin{array}{c}
  #1\\
  #2
  \end{array}
  \right)
  }
  
\title{A nonBayesian view of Hempel's paradox of the ravens}
\author{Yudi Pawitan\\
Department of Medical Epidemiology and Biostatistics\\
Karolinska Institutet\\
yudi.pawitan@ki.se}

\begin{document}
\maketitle

\begin{abstract}
In Hempel's paradox of the ravens, seeing a red pencil is considered as supporting evidence that all ravens are black. Also known as the Paradox of Confirmation, the paradox and its many resolutions indicate that we cannot underestimate the logical and statistical elements needed in the assessment of evidence in support of a hypothesis. Most of the previous analyses of the paradox are within the Bayesian framework. These analyses and Hempel himself generally accept the paradoxical conclusion; it feels paradoxical supposedly because the amount of evidence is extremely small. Here I describe a nonBayesian analysis of various statistical models with an accompanying likelihood-based reasoning. The analysis shows that the paradox feels paradoxical because there are natural models where observing a red pencil has no relevance to the color of ravens. In general the value of the evidence depends crucially on the sampling scheme and on the assumption about the underlying parameters of the relevant model. 
\end{abstract}

\section{Introduction}
According to Nicod's criterion in philosophy, observing a black raven is evidence in support of the statement that all ravens are black. The criterion further states that observing a non-black non-raven is irrelevant to the statement. This seems reasonable, but Carl Hempel (1945) came up with a well-known paradoxical counter-example. First, the following two statements are logically equivalent:
\begin{quote}
    R1:\ All ravens are black.\\
    R2:\ All non-black things are non-raven.
\end{quote}
By Nicod's criterion, seeing a non-black non-raven, for instance a red pencil, is evidence for R2. Therefore, by equivalence, observing a red pencil is evidence that all ravens are black. This violates the criterion and our common sense: How can observing a red pencil tell us anything about the color of ravens? As we walk around every day, we see an enormous number of non-black non-raven objects. Intuitively, they tell us nothing about the color of ravens. Moreover, based on the same logic, observing a red pencil also provides evidence for `All ravens are white.' What is wrong with this reasoning?

The literature on the ravens paradox is surprisingly large; surprising, because you would think that resolving such a small paradox should be simple enough. Yet, philosophers have argued about it for many decades. A peek at its Wikipedia entry would give you an overview of the many proposed resolutions and a more complete set of references. It is not my aim to review the existing literature. As a paradox of confirmation, the ravens paradox is of general interest because it raises a basic question about what it means to have supporting evidence. Hempel himself accepted the paradoxical conclusion -- a red pencil provides evidence that all ravens are black -- and declared that the impression of a paradox is a `psychological illusion.' According to him, the illusion arises because we have much prior knowledge about a red pencil, so its evidential support for the non-black non-raven property is somehow lost.

Many authors -- perhaps guided by the Bayesian analyses -- in fact accept the paradoxical conclusion. The paradox occurs supposedly because the confirmatory evidence is so tiny, as there are innumerable non-black objects compared with the obviously limited number of ravens. In any case, all the theories developed so far actually assume countable objects. To avoid the innumerability issue (cf. Hintikka, 1970), let's consider old/young people and their vaccination status. Let's call a person young if his/her age is less than 40, so that there is no ambiguity in what young means. The following two statements are logically equivalent:
\begin{quote}
    S1:\ All old people are vaccinated.\\
    S2:\ All unvaccinated people are young.
\end{quote}
If observing an unvaccinated young person is supporting evidence for S2, does it also support S1? If you accept the reasoning of the ravens paradox, you must also accept here that seeing an unvaccinated young person is evidence that all old people are vaccinated.

Most of the previous analyses of the paradox are within the Bayesian framework. The only nonBayesian analysis I know is by Royall (1997, Appendix). My attempt here is an extension of Royall's by providing additional statistical models and the accompanying likelihood-based reasoning. The key result is that the paradox feels paradoxical because there are natural models where observing a red pencil has no relevance to the color of ravens. Explicit technical-statistical assumptions are needed for the paradoxical conclusion to hold.

\section{Data, models and their likelihood}

To make our reasoning precise, we shall express it in a standard statistical framework. Viewed as a statistical problem, we wish to infer something about the vaccination status of old people based on observing an unvaccinated young person. We shall see that that is impossible without certain technical assumptions about the sampling scheme and the underlying parameters of the population. 

In the simplest model, a random sample of size $n$ is taken from a population with unknown number of people. Two binary variables are recorded $(y_1, y_2)$, such as $y_1=$ age (old-young) and $y_2=$ vaccination status (yes-no). The unknown parameters are $N_1,\ldots,N_4$, presented in Table~\ref{tab:basic}. All old people are vaccinated if and only if $N_3=0$, assuming old people exist $(N_{13}>0)$. The corresponding observed data are $(n_1,\ldots,n_4)$ in the right table ($\sum n_i = n)$. Seeing an unvaccinated young person can be modelled formally as taking a random sample of size 1 and observing $(n_1=n_2=n_3=0,n_4=1)$.

\begin{table}[h!]
\small
\hspace{.65in}
\begin{minipage}{0.4\linewidth}
 \centering
\begin{center}
\begin{tabular}{lccc}
& Old & Young & Total\\
\hline
Vax &$N_1$  &$N_2$ & $N_{12}$  \\
NoVax & $N_3$ & $N_4$ & $N_{34}$  \\
\hline
Total &$N_{13}$ & $N_{24}$ & $N$
\end{tabular}
\end{center}
\end{minipage}
\begin{minipage}{0.35\linewidth}
 \centering
\begin{center}
\begin{tabular}{lccc}
& Old & Young & Total\\
\hline
Vax &$n_1$  &$n_2$ & $n_{12}$  \\
NoVax & $n_3$ & $n_4$ & $n_{34}$  \\
\hline
Total &$n_{13}$ & $n_{24}$ & $n$
\end{tabular}
\end{center}
\end{minipage}
\caption{\em \label{tab:basic} Simplest model: in the left table $N_1,\ldots,N_4$ are the unknown population parameters representing the number of people in the categories defined by age and vaccination status. A random sample of size $n$ is taken from the population. The right table shows the observed number of individuals $(n_1,\ldots,n_4)$ in the sample.}
\end{table}

\subsubsection*{Hypergeometric model}
Theoretically, according to the likelihood principle, evidence in the data is captured by the likelihood function (e.g., Pawitan, 2001, Chapter 7).
On observing $(n_1,\ldots,n_4)$, the likelihood is based on the hypergeometric probability:
$$
L(N_1,\ldots,N_4)=\frac{\combi{N_1}{n_1}\cdots \combi{N_4}{n_4}}{\combi{N}{n}}.
$$
Specifically, for $(n_1=n_2=n_3=0,n_4=1)$, the likelihood is
\begin{equation}
L(N_1,\ldots,N_4)=N_4/N.\label{eq:N4}
\end{equation}
Clearly, the likelihood gives no information about relative proportion of $N_1$ and $N_3$. The MLE is $(\widehat{N}_1=\widehat{N}_2=\widehat{N}_3=0, \widehat{N}_4=1$), at which point the likelihood is equal 1 and zero everywhere else.  So, observing an  unvaccinated young person is not evidence for all old people being vaccinated.  

It is intuitively clear that, without further assumptions, the likelihood (\ref{eq:N4}) is not informative about the  parameters $N_1$ and $N_3$. Imagine a box containing balls for 4 different colors and we try to estimate the number of balls of each color based on a random sample of size $n$. Seeing one color tells us nothing about the other colors. (This is a well-known problem of estimating species abundance; we can perform, for example, a capture-recapture study in order to get a subset of the population with known number of subjects.)

\begin{table}[h!]
\small
\hspace{.75in}
\begin{minipage}[b]{0.35\linewidth}
\centering
\begin{tabular}{lcc}
(P) & Old & Young \\
\hline
Vax &$\lambda_1$     &$\lambda_2$    \\
NoVax & $\lambda_3$ & $\lambda_4$  \\
\hline
&&
\end{tabular}
\end{minipage}
\begin{minipage}{0.35\linewidth}
 \centering
\begin{center}
\begin{tabular}{lccc}
(M) & Old & Young & Total\\
\hline
Vax &$\theta_1$     &$\theta_2$ & $\theta_{12}$  \\
NoVax & $\theta_3$ & $\theta_4$ & $\theta_{34}$  \\
\hline
Total &$\theta_{13}$ &$\theta_{24}$  & 1.0
\end{tabular}
\end{center}
\end{minipage}
\caption{\em \label{tab:models} Two possible models and their parameterizations: (P) The number of individuals in each table entry is independent Poisson with means $\lambda_1,\ldots,\lambda_4$. (M) The observed number of sampled individuals in each cell of the table is multinomial with parameters $n$ and  $(\theta_1,\ldots,\theta_4)$.}
\end{table}

\subsubsection*{Poisson model}

The Poisson model offers a more transparent view of the problem. Let's assume that the number of people in the sample $n_i$ is independent Poisson with mean $c\lambda_i, i=1,\ldots,4,$ where $\lambda_i$ is the unknown population number and $c$ is the sampling ratio; see Table~\ref{tab:models}. The sample size is $n=c\sum_i \lambda_i$. The statements can be expressed in terms of the parameters: S1: $\lambda_1>\lambda_3=0$ and S2: $\lambda_4>\lambda_3=0$.

On observing $(n_1,\ldots,n_4)$ the likelihood is 
$$
L(\lambda_1,\ldots,\lambda_4)= \prod_{i=1}^4 P(Y_i=n_i),
$$
where $Y_i$ is Poisson with mean $c\lambda_i$. It is clear that the observed number of unvaccinated young people $n_4$ carries no information about the unknown number of vaccinated and unvaccinated old folks ($\lambda_1$ and $\lambda_3$). The MLEs are 
$$
\widehat{\lambda}_i = n_i/c.
$$
If we have no idea of the total population size, then we do not know what $c$ is, and the sample is not informative about the absolute size of $\lambda_i$s. (Though it carries information about ratios of $\lambda_i$s, captured in a multinomial model.)

\subsubsection*{Multinomial models under various scenario}

For large $N_i$s the hypergeometric model is well approximated by the multinomial with sample size $n$ and probabilities $(\theta_1,\ldots,\theta_4)$, where $\theta_i = N_i/N$; see Table~\ref{tab:models}. Similarly, conditional on the sample size, the Poisson model becomes a multinomial model with the same parameters $n$ and $(\theta_1,\ldots,\theta_4)$, where $\theta_i = \lambda_i/\sum \lambda_j$. Overall, the multinomial model is both convenient and transparent, and we shall use it from hereon. In terms of $\theta$s, the statements can be expressed thus: S1: $\theta_1>\theta_3=0$ and S2: $\theta_4>\theta_3=0$.

But first there is a crucial technical issue: How do we end up with the observation of an unvaccinated young person? Consider several possibilities depending on how we sample: 
\begin{quote}
A. Take a random person from the total population list and ascertain their age and vaccination status. \\
B. Take a random person from the list of young people and ascertain their vaccination status.\\
C. Take a random sample from the list of unvaccinated people and ascertain their age. 
\end{quote}
Unlike the ravens paradox setup, exact age and vaccination status are not visible qualities, so logistically the plans are distinct from each other. When suggesting the paradox, Hempel was perhaps not thinking about sampling plan, but since these sampling plans have distinct impact on the conclusion, let's consider them carefully. We discuss later which plan is most natural to model anecdotal observations such as black ravens or red pencils.

It is also perhaps not immediately obvious that some prior knowledge of the proportions $\theta_1,\ldots,\theta_4$ will have impact on our reasoning and conclusion. For example, suppose that the proportions of young people that are vaccinated and not vaccinated are known (say $\theta_2=0.05$ and $\theta_4=0.45$, with exact values being unimportant), but  the corresponding proportions of the old ($\theta_1$ and $\theta_3$) are not known. Then, intuitively, observing an unvaccinated young person will carry no information about the vaccination-status of the old. (This is theoretically clear in the following subsections.) So, explicit assumptions about what proportions are known are necessary for a definite solution.

\subsubsection*{Sampling plan A, completely unknown proportions}
First consider the case where we assume ignorance of both the proportions of old people and vaccinated people. In other words, both margins in Table~\ref{tab:models}(M) are free. The parameters are only constrained by $(\theta_1+\theta_2+\theta_3+\theta_4)=1$; so, we have three free parameters. On observing an unvaccinated young person $(n_1=n_2=n_3=0,n_4=1)$, the likelihood is based on the multinomial probability: 
$$
L(\theta_1,\theta_2,\theta_3,\theta_4) =\theta_1^0\theta_2^0\theta_3^0\theta_4^1=\theta_4,
$$
Clearly the likelihood contains no information about the relative proportions of vaccination in the old (the parameters $\theta_1$ and $\theta_3$). The MLE is $\that_4=1$, which implies $\that_1=\that_2=\that_3=0$. Hence, if seeing an arbitrary unvaccinated young person is taken formally as a random observation from the population where nothing is known about the age and vaccination status, \emph{the observation is not evidence in support of S1.}  

\subsubsection*{Sampling plan A, known age distribution}

Now we start with the assumption that there is an equal proportion of old and young people, but we know nothing about their vaccination status; see Table~\ref{tab:prop}. On observing $(n_1=n_2=n_3=0,n_4=1)$, the likelihood is the same as before
$$
L(\theta_1,\theta_2,\theta_3,\theta_4) =\theta_1^0\theta_2^0\theta_3^0\theta_4^1=\theta_4,
$$
in the parameter space constrained by $\theta_1+\theta_3=0.5$ and $\theta_2+\theta_4=0.5$; the space has two free parameters, say $\theta_3$ and $\theta_4$. Again the likelihood is non-informative about the proportion of vaccinated and unvaccinated old people. 

\begin{table}[h!]
\begin{center}
\begin{tabular}{lccc}
& Old & Young & Total\\
\hline
Vax &$\theta_1$     &$\theta_2$ & $\theta_{12}$  \\
NoVax & $\theta_3$ & $\theta_4$ & $\theta_{34}$\\
\hline
Total &0.5 & 0.5 & 1.0
\end{tabular}
\end{center}
\caption{\em\label{tab:prop} The parameters $\theta_1$ and  $\theta_2$ represent the proportions of old and young people, and similarly $\theta_3$ and $\theta_4$, for vaccinated and unvaccinated people respectively. By assumption, $\theta_1+\theta_3=0.5$ and $\theta_2+\theta_4=0.5$.}
\end{table}

Note that the argument \emph{does not} depend on the assumption that old and young people have equal proportions; any proportions will do. Seeing an unvaccinated young person only tells us that there are some unvaccinated people ($\theta_3+\theta_4>0$). In fact, the MLE here is $\that_4=0.5$, which is of course not very precise, but the imprecision does not affect our argument. The constraint implies $\that_2=0$, i.e., all young people are unvaccinated. Thus, seeing an unvaccinated young person is evidence that all young people are unvaccinated, but this is not equivalent to S2.

\subsubsection*{Sampling plan A, known age and vaccination-status distributions}

Suppose we have information on both the proportions of old people and of vaccinated people. Table~\ref{tab:prop2} shows two possible sets of proportions. 

\begin{table}[h!]
\begin{minipage}[b]{0.50\linewidth}
\centering
\begin{tabular}{lccc}
(T1) & Old & Young & Total\\
\hline
Vax &$\theta_1$     &$\theta_2$ & 0.2  \\
NoVax & $\theta_3$ & $\theta_4$ & 0.8\\
\hline
Total &0.5 & 0.5 & 1.0
\end{tabular}
\end{minipage}
\begin{minipage}{0.50\linewidth}
 \centering
\begin{center}
\begin{tabular}{lccc}
(T2)& Old & Young & Total\\
\hline
Vax &$\theta_1$     &$\theta_2$ & 0.6  \\
NoVax & $\theta_3$ & $\theta_4$ & 0.4\\
\hline
Total &0.5 & 0.5 & 1.0
\end{tabular}
\end{center}
\end{minipage}
\caption{\em \label{tab:prop2} Two possible sets of marginal proportions of vaccinated and unvaccinated people. Under sampling plan A, Sub-table T1 trivially falsifies S1, since $\theta_3$ must be at least $0.3$. If the proportion of vaccinated people (0.2) is less than the proportion of old people (0.5), there must be unvaccinated old people in the population. In contrast, Sub-table T2 allows evidence supporting S1. }
\end{table}

In Table~\ref{tab:prop2}(T1), the proportion of vaccinated people is 0.2 and that of unvaccinated people is 0.8. Since $\theta_3+\theta_4=0.8$, and each of $\theta_3$ or $\theta_4$ has a maximum of 0.5, therefore each must be at least 0.3. So, the statement S1 is trivially false. To make the problem non-trivial, we must assume that the proportion of vaccinated people is at least 0.5; in general, this should be at least the proportion of old people. We will take Table~\ref{tab:prop2}(T2) as an example; different proportions that satisfy the non-triviality condition will lead to the same conclusion. 

Given all the marginal constraints on $\theta_1,\ldots,\theta_4$, there is one free parameter left. For example, we can take $\theta_4$ as the free parameter; in Table~\ref{tab:prop2}(T2), $0 <\theta_4< 0.4$. On observing an unvaccinated young person, the likelihood is based on the multinomial probability as before:
$$
L(\theta_1,\theta_2,\theta_3,\theta_4) =\theta_1^0\theta_2^0\theta_3^0\theta_4^1=\theta_4,
$$
but under extra constraints $\theta_1+\theta_2=0.6$ and $\theta_3+\theta_4=0.4$. Because there is only one free parameter: as $\theta_4$ varies from 0 to 0.4, the parameter $\theta_3$ varies from 0.4 to 0. So the likelihood is now informative to each of the parameters. The MLE is $\that_4=0.4$, which implies $\that_1=0.5, \that_2=0.1$ and $\that_3=0$. In this scenario observing an unvaccinated young person \emph{is} evidential support for all old people being vaccinated. 

Incidentally this supportive scenario corresponds to the standard Bayesian resolution (e.g. Good, 1960), where a prior distribution is assumed only on the number of non-black ravens (= unvaccinated old people) while \emph{all the margins are fixed}. Perhaps not so obvious, this prior implies that there is only one free unknown parameter. (Recall that, in Bayesian inference, `free' means available for update by data). The problem is that the supposed evidence in a red-pencil is then fully dependent on the choice of prior. The Wikipedia entry on the paradox indicates that the Bayesian analyses generally support the paradoxical conclusion. Good (1960) initially wrote that the conclusion does not depend on the assumption of known marginals. He corrected it later (Good, 1961), but it seems the correction is not so well known. For instance, the Wikipedia entry described the result from his 1960 paper in great detail, but not mention the 1961 paper.

\subsubsection*{Sampling plan A: summary}
To summarize the results for plan A, which looks like a natural model for the anecdotal observation of an unvaccinated young person, the evidence generally does not support S1. Support for S1 only happens under the scenario that the proportion vaccinated and the proportion of old people are both known, and that the former is larger than the latter. This supportive scenario corresponds to the standard Bayesian resolution.

\subsubsection*{Sampling plan B}
Under plan B, sampling young people and recording their vaccine status, we have a Bernoulli event with probability $p\equiv \theta_4/(\theta_2+\theta_4)$ of seeing an unvaccinated one, so the likelihood is
$$
L(p)=p,
$$
It is clear from this that we can say something about young people's vaccine status, that some or even perhaps all of the young are unvaccinated. (In fact, the MLE $\widehat{p}=1$). However, the likelihood carries no information about $\theta_1$ or $\theta_3$, so the evidence neither supports S2 nor S1. However, unlike plan A, the conclusion is not affected by the extra information on the proportions of old peopple or vaccinated people, as given in Table~\ref{tab:prop2}.

\subsubsection*{Sampling plan C}
Finally, under plan C, among the unvaccinated people, the proportion of young people is $p=\theta_4/(\theta_3+\theta_4)$.  So $p=1$ if and only if $\theta_3=0$. Observing an unvaccinated young person is a Bernoulli event with probability $p$, so the likelihood is 
$$
L(p) = p.
$$
Now, the MLE $\widehat{p}=1$ is consistent with $\that_3=0$, meaning the evidence does support S2 and S1. As for plan B, this conclusion is not affected by the extra information about the proportions of old people or vaccinated people.

\section{What if we sample thousands}
Even when the observation is supportive, the evidence based on one unvaccinated young person feels unconvincing. However, in this thought experiment, the amount of evidence is not an issue. What matters is whether the observation is a supporting evidence or not. Yet it is instructive to imagine a much larger sample size, such as 10,000, as the evidence should accumulate enough to feel more conclusive.

So, let's assume that, for each sampling plan: (i) we take a random sample of size not 1 but 10,000, and (ii) we only observe unvaccinated young people in the sample. Computationally, we just need to take the power of 10,000 to the probabilities in the likelihood functions, so we get highly concentrated likelihoods. However, we can perhaps follow the logic more intuitively using verbal explanations only. To be nontrivial, let's assume that there are old people in the population.

Under plan A, the evidence is consistent with the situation where the only people in the target population are unvaccinated young people. This means we cannot say anything meaningful about old people. It would be absurd to suppose that had there been old people, all of them would be vaccinated, as the vaccination status then comes from the supposition, not from the data.  If we assume that old people exist in a non-trivial proportion, then the sample cannot have been taken randomly from the population, so again we cannot infer anything about the vaccination status of the old. Overall, under plan A, the observed data do not tell us anything about the old people's vaccination status. 

We can contrive another result for plan A to mimic the both-margins-known scenario.  For this we drop the assumption of only observing 10,000 unvaccinated young people. Instead, from the 10,000 sampled people, we know there are 5000 old, 5000 young, 6000 vaccinated and 4000 unvaccinated people; see Table~\ref{tab:prop2}(T2) for the theoretical version. Then any information about unvaccinated young people ($n_4$) will tell us about the vaccination status of old people ($n_1$ and $n_3$). For example, if $n_4=4000$ then $n_3=0$ and $n_1=5000$, which means all old people are vaccinated. But if $n_4<4000$ then $n_3>0$, which falsifies S1. So, in large samples, seeing unvaccinated young people does not guarantee S1.

Under plan B, we sample 10,000 from the list of  young people and ascertain that all of them are unvaccinated. We have strong evidence for the statement `All young people are unvaccinated', but that's not logically equivalent to S2. We cannot make statements about all unvaccinated people: Since we only sample young people, we do not know if there are unvaccinated old people. Consequently, the evidence supports neither S2 nor S1. 

Under plan C, we sample 10,000 from the list of unvaccinated people and ascertain that all of them happen to be young. This evidence strongly supports S2. It also tells us there are no old people among the unvaccinated people, which can only be because all old people are vaccinated. Hence, the evidence indeed supports S1. 

In summary, only under sampling plan C does the evidence of seeing 10,000 unvaccinated young people support the statement that all old people are vaccinated. 

%\subsubsection*{Sample size vs population size}
%Many analyses of the ravens paradox claim that the information from observing a black raven is more than from observing a red pencil. One commonly stated reason is that the number of non-black non-ravens is substantially more than the number of ravens. Let's think about the surveys that people do regularly, such as political or consumer surveys. Assuming the population size is large enough, so there is no finite-population effect, which one matters: the sample size or the population size? Obviously the former. A survey of 1000 people conducted in the USA (population 331M) carries the same statistical information as the one done in Ireland (population 4.9M).  Thus, if we consider the parameter $\theta_3$ in Table~\ref{tab:prop} as the target parameter. A properly conducted study based on sampling of old people and another based on sampling of unvaccinated people will carry the same information about $\theta_3$, as long as they are of the same size. 

%Still, intuitively we feel that a large number of non-black non-raven things we see in daily life does not carry evidence about the color of ravens. The main problem is of course we are not observing objects in a statistically random way; most of us are normally not even in an environment where we can observe ravens. There is a limit on the viability of the ravens in as a thought experiment. 

\section{Conclusions}
%From the OXford Dictionary of Philosophy: A condition governing the confirmation of a general hypothesis by particular pieces of evidence, proposed by the French philosopher Jean Nicod (1893–1924) in his Foundations of Geometry and Induction (1930). It requires that an instance of a generalization that all As are B provides a positive, confirming piece of evidence for the generalization; evidence of something that is neither A nor B is irrelevant to it, as is evidence of something that is B but not A. The principle is put under pressure by Hempel's paradox, which apparently yields circumstances in which something that is neither A nor B may confirm the generalization.

Our analysis shows that Hempel's paradox seems paradoxical because there are natural scenarios where observing a red pencil -- or even thousands of them -- tells us nothing about the color of ravens. Unlike Hempel or the other previous analyses, the current analysis conforms with our common sense that there is no sufficient reason to accept the paradoxical conclusion. An explicit statement on how the observation was made and some technical-statistical assumptions about the underlying parameters are needed. Perhaps obvious to scientists, support for a statement or a hypothesis requires a systematic gathering of evidence. Hempel's ravens paradox highlights the pitfalls in trying to interpret results from haphazardly collected data. It is interesting that our antennae are usually well-tuned for problems that arise with such data, but when put in a different context as in Hempel's paradox, the problem somehow turns into a serious puzzle that generates lengthy debates.

We have gone through different probability models to capture the observation that underpins the paradox. The multinomial model under different sampling plans captures the problem most conveniently and transparently. Since the different sampling plans can have different conclusions, which one is the most natural or appropriate? On first impression, sampling plan A, where we just sample from the total population, seems the most appropriate. But, under this plan, the paradoxical conclusion holds only if we assume that both of the marginal proportions are known. This assumption seems unlikely to hold as we casually move around and see myriads of objects in our environment. Plan C, where we sample from unvaccinated people and ascertain their age, is the safest plan to establish supporting evidence for S1. But this plan requires a careful phrasing/reporting of the evidence, which does not conform to the original anecdotal phrasing of the paradox. 

%How come we do not have such sampling issues when we observe direct evidence of black ravens? It seems we could just walk around seeing more and more black ravens, and get more and more convinced that all ravens are black. In our mind we are naturally but unconsciously gathering data on ravens and keeping track of their color. We can see this by our reporting (if somebody asks): we would say `all the ravens we have seen so far are black'. Statistically this does support the statement that all ravens are black. But nobody will ever report `all the non-black things we saw were non-ravens,' because there is no meaningful data gathering on the properties of `non-black things,' so we are all the time within the sampling plan A.

%\subsubsection*{Nicod's criterion vs plausible reasoning}
Hempel's paradox is a counter-example of the Nicod's criterion. Would the paradox also occur in plausible reasoning (Polya, 1954)? It is an important question because plausible reasoning is a cornerstone of scientific reasoning: we have a theory G that predicts a specific evidential instance E, i.e. a conditional statement `G implies E.' Then, ascertaining E in an experiment increases the plausibility of G. Let G be the statement `All old people are vaccinated,' and E be the statement `we take a random old person, and ascertain that he is vaccinated.' Clearly, G implies E, and in this case E does increase the plausibility of G. 

However, the statement `We take a random vaccinated person, and ascertain that he is old' is \emph{not} implied/guaranteed by G, because there could be vaccinated young person too. So, just observing a vaccinated old person without any explanation on how they came under observation does not increase the plausibility of G. The correct application of plausible reasoning in this case also requires a proper sampling plan to provide evidence for G. In other words, Nicod's criterion does not follow the rule of plausible reasoning, and Hempel's paradox is not a counter-example of plausible reasoning.

\section*{References}

\begin{description}

%\item {\sc Birnbaum, A.}\ (1962). On the foundation of statistical inference. \emph{Journal of the American Statistical Association} \textbf{57}, 269--326.

\item{\sc Good, I. J.} (1960). The Paradox of Confirmation. \emph{British Journal for the Philosophy of Science,} 11 (42): 145–149.

\item{\sc Good, I. J.} (1961). The Paradox of Confirmation (II). \emph{British Journal for the Philosophy of Science,} 12(45), 63–64. 

\item{\sc Hempel, C. G.} (1945). Studies in the Logic of Confirmation I. \emph{Mind}, 54 (13): 1–26.

\item{\sc Hintikka, J.} (1970). Inductive independence and the paradoxes of confirmation. In Rescher, Nicholas (ed.). \emph{Essays in honor of Carl G. Hempel: A tribute on the occasion of his sixty-fifth birthday.} Synthese library. Dordrecht: D. Reidel. pp. 24–46.

%\item{Maher, P.} (2004). Probability Captures the Logic of Scientific Confirmation. In Christopher Hitchcock (ed.). \emph{Contemporary Debates in the Philosophy of Science.} Blackwell. pp. 69–93.

\item  {\sc Pawitan Y.} (2001). \emph{In all likelihood:
Statistical modelling and inference using likelihood}. 
Oxford University Press, Oxford, UK.

\item {Polya, G.} (1954). \emph{Mathematics and Plausible Reasoning Volume I: Induction and Analogy in Mathematics.} Princeton University Press, Princeton, New Jersey.

\item {Royall, R.M.} (1997). \emph{Statistical Evidence: A Likelihood Paradigm.} New York: Chapman \& Hall.

\end{description}
\end{document}